\documentclass[12pt]{report}
\usepackage[dvips]{graphicx}
\newcommand{\sptwo}{1.4}

\newcommand{\doublespace}{\edef\baselinestretch{\sptwo}\Large\normalsize}

\begin{document}
\doublespace
\begin{center}
{\bf Time-Dependent Density-Functional Theory for Trapped Strongly-Interacting
Fermionic Atoms}\\
\renewcommand\thefootnote{\fnsymbol{footnote}}
{Yeong E. Kim \footnote{ e-mail:yekim$@$physics.purdue.edu} and
Alexander L. Zubarev\footnote{ e-mail: zubareva$@$physics.purdue.edu}}\\
Purdue Nuclear and Many-Body Theory Group (PNMBTG)\\
Department of Physics, Purdue University\\
West Lafayette, Indiana  47907\\
\end{center}

\begin{quote}
The dynamics of strongly interacting trapped dilute Fermi gases (dilute in the sense that
 the range of interatomic potential is small
compared with inter-particle spacing
) is investigated in a single-equation approach to the time-dependent 
density-functional theory. 
Our results are in good agreement with recent experimental data in the BCS-BEC
crossover regime.

It is also shown  that
%, in regimes which may now accessible experimentally,
the calculated corrections to the  hydrodynamic 
approximation may be important even for systems with a rather large number of
 atoms.
\end{quote}

\vspace{5mm}

\pagebreak
The recently reported ultracold trapped Fermi gases with tunable atomic
 scattering length [1-11] in the vicinity of a Feshbach resonance stimulated a
 large number of theoretical investigations. Some of these works are  based on
 the
 assumption that the properties of  strongly interacting dilute Fermi gas
 at zero temperature are well described by the  hydrodynamic
 approximation  (HA) [12-15]
$$
\frac{\partial n}{\partial t}+\nabla (n \vec{\mbox{v}})=0,
\eqno{(1)}
$$

$$
\frac{\partial \vec{\mbox{v}}}{\partial t}+\frac{1}{m}\nabla (V_{ext}+\frac{
\partial
(n \epsilon(n))}{\partial n}+\frac{1}{2} m \mbox{v}^2)=0,
\eqno{(2)}
$$
where $n$ is the density, $\epsilon(n)$ is the ground-state energy per particle
 of the homogeneous system and $\vec{\mbox{v}}$ is the velocity field.

In this paper
the dynamics of strongly interacting trapped dilute Fermi gases (dilute in the 
sense that
 the range of interatomic potential is small
compared with inter-particle spacing
) is investigated in the single equation approach to the time-dependent
density-functional theory covering the whole crossover region at zero
 temperature.  
It is shown 
%, in regimes
 %now accessible  experimentally,
 for the case of elongated cigar-shaped harmonic
traps that
the calculated corrections to the HA
 may be important even for systems  with a rather large number of
 atoms.

We mention here  Refs.[16] where  an extension of the density-functional theory
(DFT) to superconducting systems [17] was generalized to a number of nuclear and
atomic systems.

 Let us consider a Fermi gas consisting  of a 50-50 mixture of two different states confined in a harmonic trap $V_{ext}(\vec{r})=(m/2)(\omega_{\perp}^2 (x^2+y^2)+
\omega_z^2 z^2)$.
% The S-wave scattering length between the two fermionic species is assumed to be negative, $a<0$. 
In Eq.(2), the kinetic-energy density $t(n)$ is approximated by the Thomas-Fermi (TF) kinetic-energy density $t_{TF}(n)=(3/10) n \hbar^2 k_F^2/m$, where
$k_F=(3 \pi^2 n)^{1/3}$.
For slowly varying densities characterized by the condition 
$\mid \nabla n\mid/n^{4/3}\ll1$, the kinetic energy density is well 
represented by the Kirzhnitz gradient expansion (KGE) [18]
$t(n)=t_{TF}(n)+t_W(n)/9+...$,
where $t_W(n)=(\hbar^2/(8 m))(\nabla n)^2/n$ is the original von Weizs\"{a}cker
density  (OWD)[19], which gives the entire kinetic energy density of noninteracting bosons.

%For $\lambda=\omega_z/\omega_{\perp}\approx 10^{-2}$ and a very large number of atoms, $N \geq 10^5$ the ratio  $\mid \nabla n\mid/n^{4/3}$ is small
%at equilibrium and the 
%Kirzhnitz correction is negligible, but for $N \leq 10^4$, and $\lambda N>1$
%the convergence criteria of the KGE at equilibrium is violated. Remembering
% that $t_W(n)$ gives the entire kinetic energy density of noninteracting bosons, the straightforward estimate of a correction to $t_{TF}(n)$ in this case is
% provided by $t_W(n)$.
In the case of large but finite number of atoms $N$, the density $n$ is not constant. At small distances the ratio  $\mid \nabla n\mid/n^{4/3}$ is small and both the Kirzhnitz correction and the OWD are negligible. 
On the contrary, near the surface the Hartree-Fock (HF) type densities are
 proportional to the square of the last occupied state. Therefore, the OWD is
 important in this case and it is expected to determine the asymptotic
 behavior of the density at large distances. It is also expected that the OWD is important in the case of the tight radial trapping, $\lambda \ll 1$.
  In Refs.[20], the OWD
 was considered as a correction to the TF kinetic-energy density.

Adding the OWD to $t_{TF}(n)$  we have
$$
\frac{\partial \vec{\mbox{v}}}{\partial t}+\frac{1}{m}\nabla (V_{ext}+\frac{
\partial(n \epsilon(
n))}{\partial n}+\frac{1}{2} m \mbox{v}^2-\frac{\hbar^2}{2 m} \frac{1}{\sqrt{n}}
 \nabla^2 \sqrt{n})=0.\eqno{(3)}
$$
We define the density of the system as
$
n(\vec{r},t)=\mid \Psi(\vec{r},t)\mid^2,
$
and the velocity field $\vec{\mbox{v}}$ as
$
\vec{\mbox{v}}(\vec{r},t)=\hbar(\Psi^\ast \nabla \Psi-
\Psi \nabla \Psi^\ast)/(2 i m n(\vec{r},t)).
$
From  Eqs.(1) and (3), we obtain  the following  
nonlinear Schr\"{o}dinger equation
$$
i\hbar \frac{\partial \Psi}{\partial t}=-\frac{ \hbar^2}{2 m} \nabla^2 \Psi
+V_{ext} \Psi+\frac{\partial( n \epsilon(n))}{\partial n}\Psi,
\eqno{(4)}
$$
 which is equivalent, to a certain extent,  to the single equation approach of 
Deb et al. [21] to
 the time-dependent
density-functional theory (TDDFT). 

If the trap potential, $V_{ext}$, is independent of time, one can write 
 $\Psi(\vec{r},t)=\Phi(\vec{r}) \exp(-i \mu t/\hbar
)$,
where $\mu$ is the chemical potential, and $\Phi$ is normalized to the total 
number of particles, $\int d\vec{r} \mid \Phi \mid^2=N$. Then Eq.(4) becomes 
$$
(-\frac{ \hbar^2}{2 m} \nabla^2 +V_{ext}+\frac{\partial( n \epsilon(n))}
{\partial n})\Phi=\mu \Phi,
\eqno{(5)}
$$
where the solution of the equation (5) minimizes
the energy functional
$
E=N<\Phi\mid -\frac{ \hbar^2}{2 m} \nabla^2+V_{ext}+\epsilon(n)\mid \Phi>,
$
and the chemical potential $\mu$ is given by $\mu=\partial E/\partial N$.

In order to take into account atoms lost  by inelastic collisions,
 we
 model the loss by the rate equation
$$
\frac{d N}{d t}=-\int \chi(\vec{r},t) d\vec{r},
$$
where $\chi(\vec{r},t)=\sum_{l=1} k_l n^l g_l(n)$,  $n^l g_l$ is the local
$l$-particle correlation function and $k_l$ is the rate constant for the
$l$-body atoms loss.
The generalization of Eq.(4) for the case of inelastic collisions reads [22]
$$
i\hbar \frac{\partial \Psi}{\partial t}=-\frac{ \hbar^2}{2 m} \nabla^2 \Psi
+V_{ext} \Psi+\frac{\partial( n \epsilon(n))}{\partial n}\Psi- i \frac{\hbar}{2}
 \sum_{l=1} k_l n^{l-1} g_l(n)\Psi.
\eqno{(6)}
$$
For the negative S-wave scattering length between the two fermionic species,
 $a<0$,
in the low-density regime, $k_F\mid a \mid \ll 1$, the ground state energy per
particle , $\epsilon(n)$, is well represented by an expansion in power of
$k_F \mid a \mid$ [26]
$$
\epsilon(n)=2 E_F[\frac{3}{10}-
\frac{1}{3 \pi} k_F \mid a \mid+0.055661 (k_F\mid a \mid)^2
-0.00914 (k_F\mid a \mid)^3+...],
\eqno{(7)}
$$
where $E_F=\hbar^2 k_F^2/(2 m)$.
In the opposite regime, $a\rightarrow - \infty$
(the Bertsch many-body problem, quoted in
Refs.[27]), $\epsilon(n)$ is proportional to that of the non-interacting Fermi
 gas
$$
\epsilon(n)=(1+\beta)\frac{3}{10} \frac{\hbar^2 k_F^2}{m},
\eqno{(8)}
$$
where a  universal parameter  $\beta$  is estimated to be $\beta=-0.56$ [28].

 In the $a\rightarrow +0$ limit the system reduces to the
dilute Bose gas of dimers [29]
$$\epsilon(n)=E_F(-1/(k_F a)^2+a_m k_F/(6 \pi)+...,
)
\eqno{(9)}
$$
where $a_m$ is the boson-boson scattering length.
While the BCS mean-field theories [30] predict  $a_m=2a$ [31],
a solution of 4-fermion problem for contact scattering provided the value 
$a_m\approx 0.6 a$ [32].

Very little is known about the correct form of $\epsilon(n)$ in the 
intermediate range. Therefore, a simple interpolation of the form 
$\epsilon(n)\approx 
E_F P(
k_F  a)$ with a smooth function $P(x)$ mediating between the known limits 
suggests itself as a pragmatic alternative.

In Ref.[33] it has  been proposed a [2/2] Pad\'{e} approximant for the function $P(x)$ for the negative $a$
$$
P(x)=\frac{3}{5}-2\frac{\delta_1\mid x \mid+\delta_2 x^2
}{1+\delta_3\mid x\mid+\delta_4 x^2},
\eqno{(10)}
$$
where $\delta_1=0.106103$, $\delta_2=0.187515$, $\delta_3=2.29188$,      
$\delta_4=1.11616$.
Eq.(10) is constructed to reproduce the first four terms of the expansion (6) in
 the low-density regime and
 also to  reproduce
exactly results of
 the recent Monte Carlo calculations [28], $\beta=-0.56$, in the  unitary limit,
$k_F a \rightarrow -\infty$. 
 
 For the positive $a$ case ( the interaction is strong enough to form bound molecules with energy $E_{mol}$)  we consider a [2/2] Pad\'{e} approximant
$$
P(x)=\frac{E_{mol}}{2 E_F}+\frac{\alpha_1 x+\alpha_2 x^2}{1+\alpha_3 x+\alpha_4 x^2},
\eqno{(11)}
$$
where 
 parameters $\alpha$ are fixed by two continuity conditions at large $x$,
$1/x\rightarrow 0$, and by two continuity conditions at small $x$. 
For example, $\alpha_1=0.0316621$, $\alpha_2=0.0111816$, $\alpha_3=0.200149$,
and $\alpha_4=0.0423545$ for $a_m=0.6 a$.

Fig. 1 and Fig. 2 show the comparison between  [2/2] Pad\'{e} approximations,
Eqs.(10,11),  
and the  lowest order constrained variational (LOCV) approximation [34] and
 the BCS mean-field theory for 
$\epsilon(n)$.
The LOCV calculations  agree very well with the [2/2] Pade
 approximation results on the BCS side ($a<0$). It is evident the difference
 between  our results and the BCS mean-field theory calculations. For example,
the BCS mean-field gives $\beta=-0.41$. We mention here that
 $\epsilon(n)/E_F$ on the BCS side  ($a<0$) and $(\epsilon(n)+|E_{mol}|/2)/E_F$
 on the BEC side ($a>0$) show a
 smooth monotonic behavior as a function of $k_Fa$.

The predictions of Eq.(5) with $\epsilon(n)$ from Eq.(10) for the axial cloud
 size of   strongly interacting $^6Li$ atoms are shown in Fig 3  [35]. It 
indicates
that the TF approximation of the kinetic energy density is a very good 
approximation for the experimental conditions of Ref.[11], $N \lambda \approx 
10^4$  (inclusion of
 the OWD gives a negligible effect, $<0.5\%$) [37]. 

It can be proved [24] that every solution of  equation (4)
is a stationary point of an action corresponding to the Lagrangian density
$$
\mathcal{L}_0=\frac{i\hbar}{2}(\Psi\frac{\partial\Psi^{\ast}}{\partial t}-
\Psi^{\ast}\frac{\partial\Psi}{\partial t})+\frac{\hbar^2}{2m}\mid \nabla \Psi
\mid^2+\epsilon(n)n+V_{ext}n,
$$
which for $\Psi=e^{i \phi(\vec{r},t)} n^{1/2}(\vec{r},t)$ can be rewritten as
$$
\mathcal{L}_0=\hbar \dot{\phi}n+\frac{\hbar^2}{2m} (\nabla \sqrt{n})^2+
\frac{\hbar^2}{2m}n (\nabla \phi)^2+\epsilon(n)n+V_{ext}n.
$$
For a time-dependent harmonic trap, $V_{ext}(\vec{r},t)=
(m/2)\sum_{i=1}^3 \omega_i^2(t) x_i^2$, a suitable trial function can be taken
 as
$\phi(\vec{r},t)=\chi(t)+(m/(2 \hbar) \sum_{i=1}^3 \eta_i(t) x_i^2$,
$n(\vec{r},t)=n_0(x_i/b_i(t))/\zeta(t)$, where $\zeta(t)=\prod_jb_j$.
With this ansatz, the Hamilton principle, $\delta\int dt\int\mathcal{L}_0 d^3 r=
0$, gives the following equations for the scaling parameters $b_i$
$$
\ddot{b}_i+\omega_i^2(t) b_i-\frac{2 <T_i>}{m <x_i^2> b_i^3}-
\frac{1}{m <x_i^2> b_i}
\int[n^2d\epsilon(n)/dn]_{n=n_0(\vec{r})/\zeta(t)} d^3 r 
\zeta(t)=0,
\eqno{(12)}
$$
where $b_i(0)=1$, $\dot{b}_i(0)=0$ and $\omega_i=\omega_i(0)$ fix the initial
configuration of the system, corresponding to the density $n_0(\vec{r})$ and
$<T_i>=-\hbar^2/(2 m N) \int n^{1/2}(\partial^2/\partial x_i^2) n^{1/2} d^3 r$,
$<x_i^2>=(1/N) \int  n x_i^2 d^3 r$.

Expanding Eqs.(12) around equilibrium ($b_i=1$) we get the following equations 
for the collective frequencies, $\omega$
$$
(2 +\kappa_i -\frac{\omega^2}{\omega_i^2}) y_i+(1+\frac{1}{2} \kappa_i+\chi_i)
 (y_1+y_2+y_3)=0,
\eqno{(13)}
$$
where $\kappa_i=4 <T_i>/(m \omega_i^2 <x_i^2>)$ and $\chi_i=\int n_0^3 
\partial^2 \epsilon/(\partial n_0^2) d^3 r/(m \omega_i^2 <x_i^2>)$.

In Table I we give the calculated values of the radial breathing mode
frequency, $\nu=\omega_{rad}/(2 \pi)$, of highly degenerate gas of $^6Li$
 atoms near a Feshbach resonance at 822 G [41].
It can be seen from Table I, that the difference between two approximations,
 Eqs.(1,2) and Eq.(4), is less than 0.75\%, and both approximations give a very
good agreement with experimental data of Ref.[1]. The parameter $\lambda N$ 
for this case is very large, $\lambda N\geq 10^4$.

In Fig. 4, we present the calculations for the frequency of the radial 
compression mode $\omega_{rad}$ as a
function
 of the dimensional parameter $(N^{1/6} a/a_{ho})^{-1}$ in the case of an anisotropic trap ($\omega_x=\omega_y=\omega_{\perp}$,
$\omega_z/\omega_{\perp}=\lambda$). One can easily see that the corrections to
the
  hydrodynamic
approximation
(HA), Eqs.(1)
 and (2), are important 
%is not valid
 even
 for relatively large $N$ and $\lambda N$. For example,
the correction to  $\omega_{rad}$ in unitary limit is larger than 11\% and 25\% for $\lambda=10^{-2}$, $N=10^4$ and $\lambda=10^{-2}$, $N=10^3$,
 respectively. 

In the HA, $\omega_{rad}$ is independent of $N$ for a fixed  
$(N^{1/6} a/a_{ho})^{-1}$. The deviation from this behavior does not 
demonstrate the cross-over to the $1D$ behavior, since $\lambda N>1$ [42].
% (the transverse motion is ``frozen" at zero temperature if $N\lambda \ll 1$
% [36]).
It demonstrates that the validity of the HA depends on the properties of the
 trap. In Ref.[43] it was shown that, for the case of isotropic trap,
 $\lambda=1$, with $N=20$ and $N=240$, the TF approximation reproduces the energy within accuracies of 2\%  and  1\%,  respectively.

In Fig.5, the calculated radial compressional frequency is compared with experimental data [1] in the BCS-BEC crossover region. There is a very good agreement
 between  calculations and experimental data [1]. However our calculations for $\omega_{rad}$ disagree with experimental data of Ref.[44].

In the present paper, we have used Eq.(4). The next step is to develop the
Kohn-Sham time-dependent DFT [45]  for
   two-component Fermi
 gases in elongated traps ($\lambda \ll 1$),
  which we will
 consider in our future work.

In conclusion, 
the dynamics of strongly interacting trapped dilute Fermi gases 
%(dilute in the
%sense that
% the range of interatomic potential is small
%compared with inter-particle spacing)
 is investigated in the single equation approach to the time-dependent
density-functional theory. 
Our results are in good agreement with recent experimental data in the BCS-BEC
crossover regime.

It is also shown  that
the calculated corrections to the  hydrodynamic
approximation may be  important even for systems with a rather large number of
 atoms.

ALZ thanks V. Dunjko for useful discussion.

{\it Note added.} - While this work was being prepared for publication, 
preprints [46] appeared in which the authors calculate the equation of state,
 $\epsilon(n)$, using the quantum  Monte Carlo method. Their results are in a
good  agreement with ours Pad\'{e} [2/2] approximation for both negative and
 positive scattering length.
\pagebreak

Table I. Radial breathing mode frequencies $\nu=\omega_{rad}/(2 \pi)$ of highly
 degenerate gas of $^6Li$ atoms near a 822(G) Feshbach resonance [36].
$B$ is applied magnetic field, $\nu_{exp}$ indicate experimental data from the
 Duke University group [1], $\nu$ and $\nu_{TF}$ represent theoretical 
calculations
 that use equation (4) and the hydrodynamic approximation (equations (1) and
(2)), respectively.  The trap parameters are $\omega_{\perp}=2 \pi \times 1549$, $\omega_z=2 \pi \times 70$.

\vspace{8pt}
\begin{tabular}{lllll}
\hline\hline
$B(G)$
&$N(10^3)$
&$\nu_{exp}$(Hz) [1]
&$\nu$(Hz)
&$\nu_{TF}$(Hz) \\ \hline
860
&294
&2857
&2810
&2793 \\ \hline
870
&288
&2837
&2804
&2787 \\ \hline
870
&225
&2838
&2806
&2786 \\ \hline
870
&379
&2754
&2803
&2788 \\ \hline
870
&290
&2775
&2804
&2787 \\ \hline
870
&244
&2779
&2805
&2787 \\ \hline
880
&258
&2836
&2800
&2783 \\ \hline
910
&268
&2798
&2792
&2775
\\ \hline\hline
\end{tabular}

\pagebreak
\begin{figure}[ht]
\includegraphics{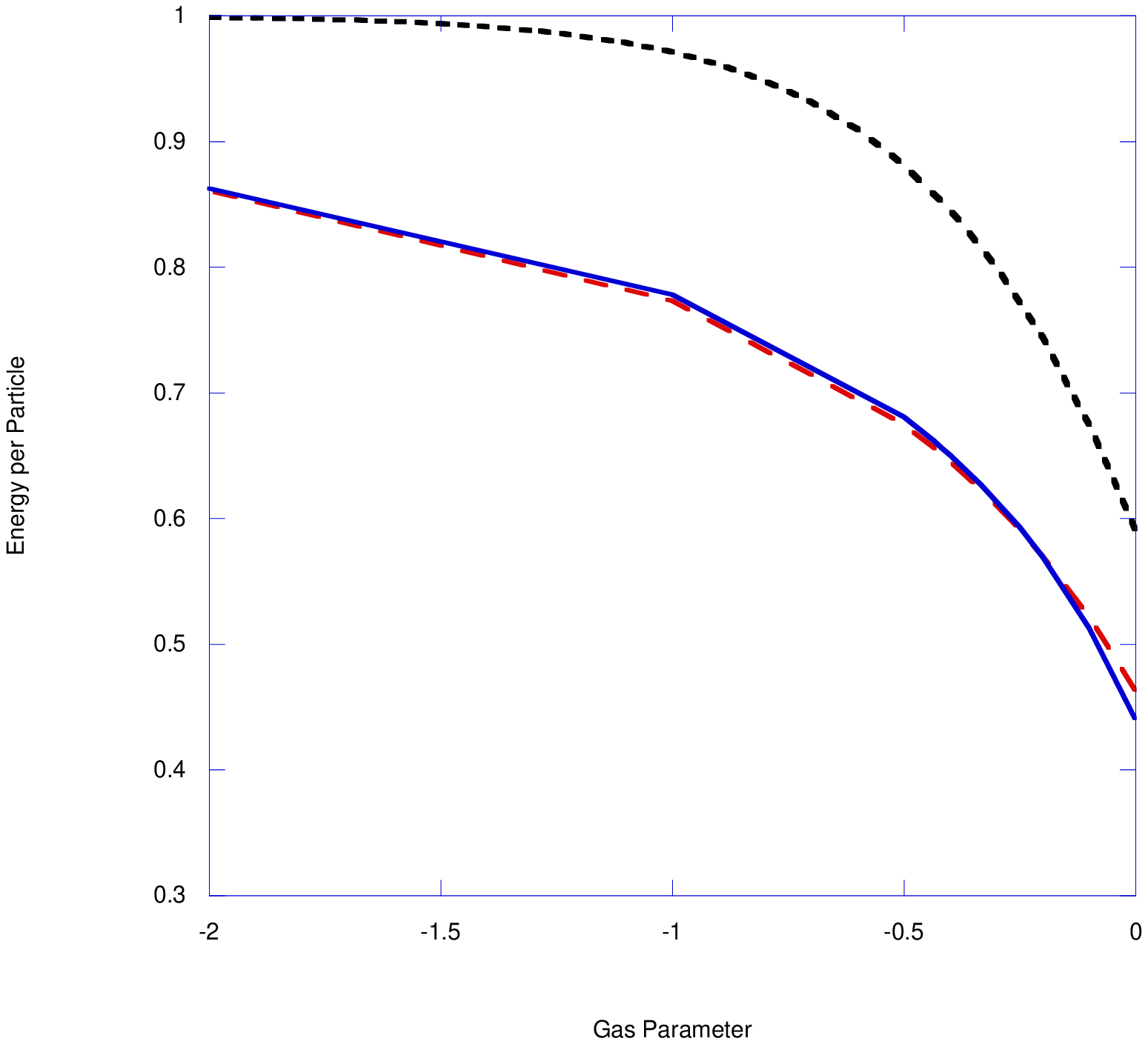}
\end{figure}
Fig.1. The ground state energy per particle, $\epsilon(n)$,
  in units of $3 \hbar^2 k_F^2/(10 m)$ as a function of the gas  parameter
 $(k_F a)^{-1}$. The solid line,
the long dashed line  and the short dashed line
 represent the results calculated
using the [2/2] Pad\'{e} approximation, Eq.(10), the LOCV approximation,
 and the BCS mean-field theory,  respectively.

\pagebreak
\begin{figure}[ht]
\includegraphics{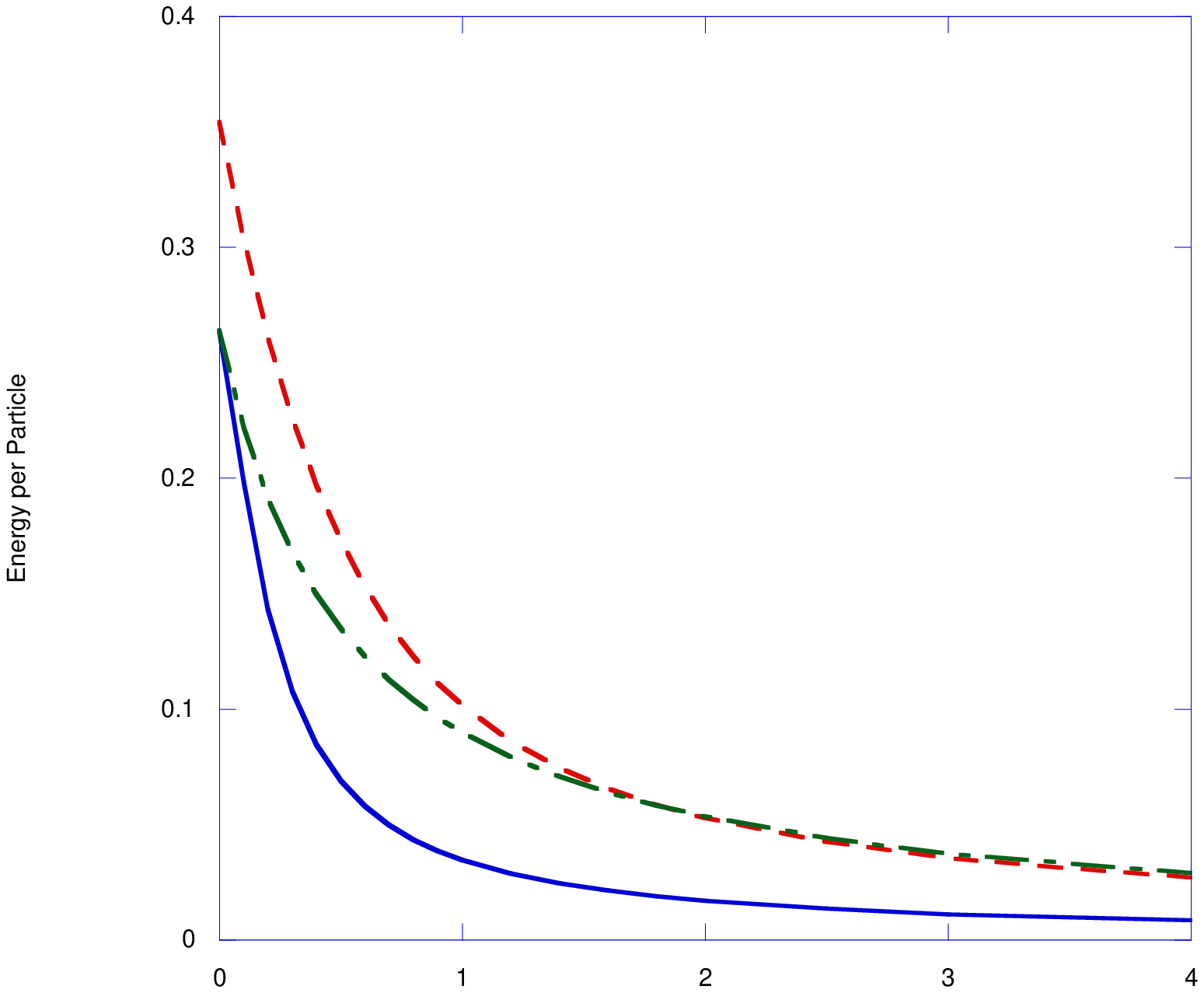}
\end{figure}
\begin{center}
$(k_F a)^{-1}$
\end{center}
Fig.2. The ground state energy per particle, $\epsilon(n)+|E_{mol}|/2$,
  in units of $\hbar^2 k_F^2/(2 m)$ as a function of the gas parameter
 $(k_F a)^{-1}$.
The dashed line,  the dotted-dashed line and
the solid line
represent the results calculated using the BCS mean-field theory, the [2/2]
Pad\'{e}
 approximation, Eq.(11), with $a_m=2a$, and $a_m=0.6a,$ respectively.

\pagebreak

\begin{figure}[ht]
\includegraphics{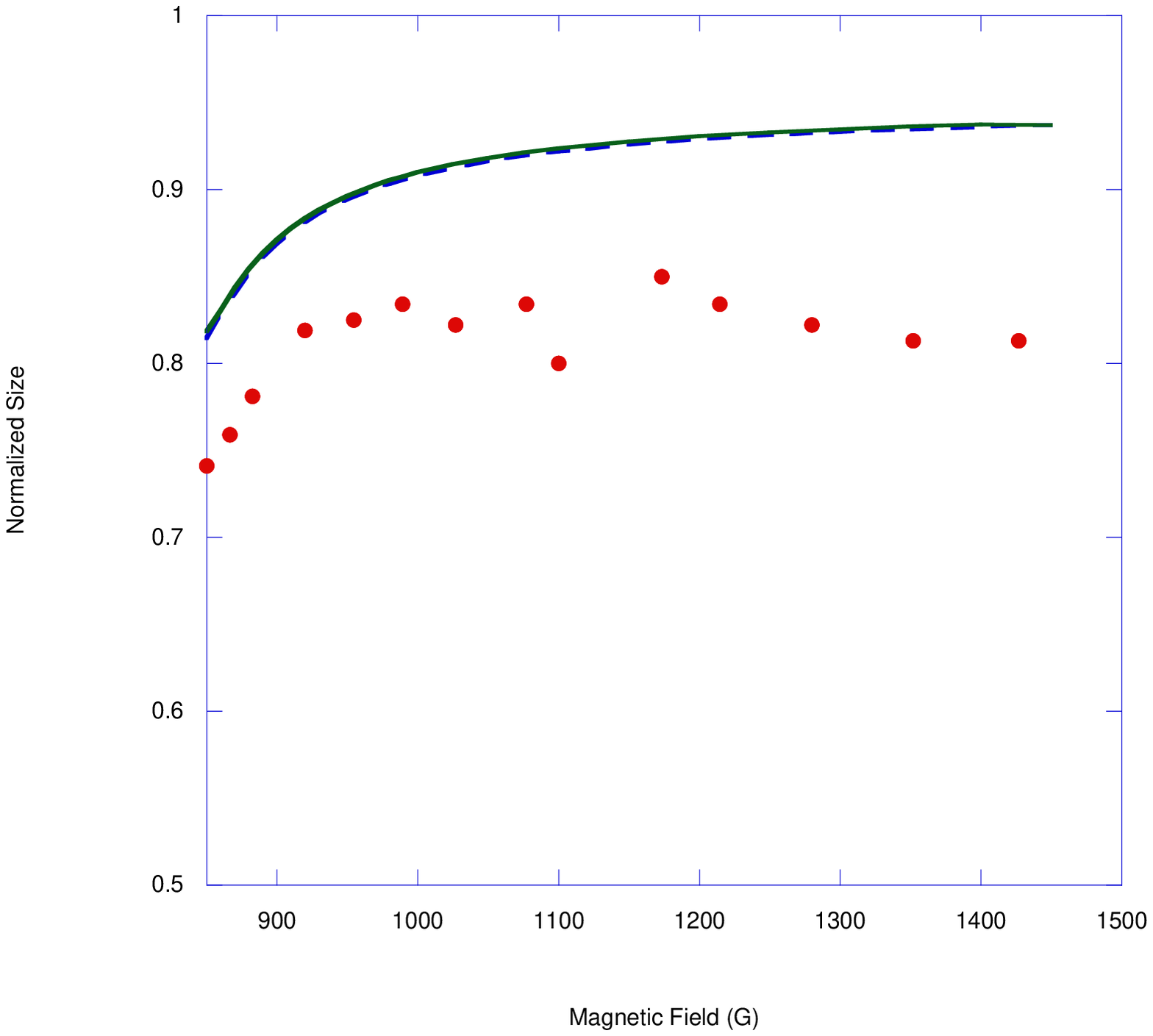}
\end{figure}
Fig. 3. Axial cloud size of strongly interacting $^6Li$ atoms
 after normalization to a non-interacting Fermi gas
with $N=4 \times 10^5$ atoms as a function of the magnetic field $B$ [32].
 The trap parameters are $\omega_{\perp}=
2 \pi \times 640$Hz, $\omega_z=2 \pi (600 B/kG+32)^{1/2}$Hz.
 The solid line and dashed line  represent the results of
 theoretical calculation that includes the OWD or uses the TF approximation
 for the kinetic energy density, respectively. The circular dots indicate 
 experimental data from the Innsbruck group [11].

\pagebreak

\begin{figure}[ht]
\includegraphics{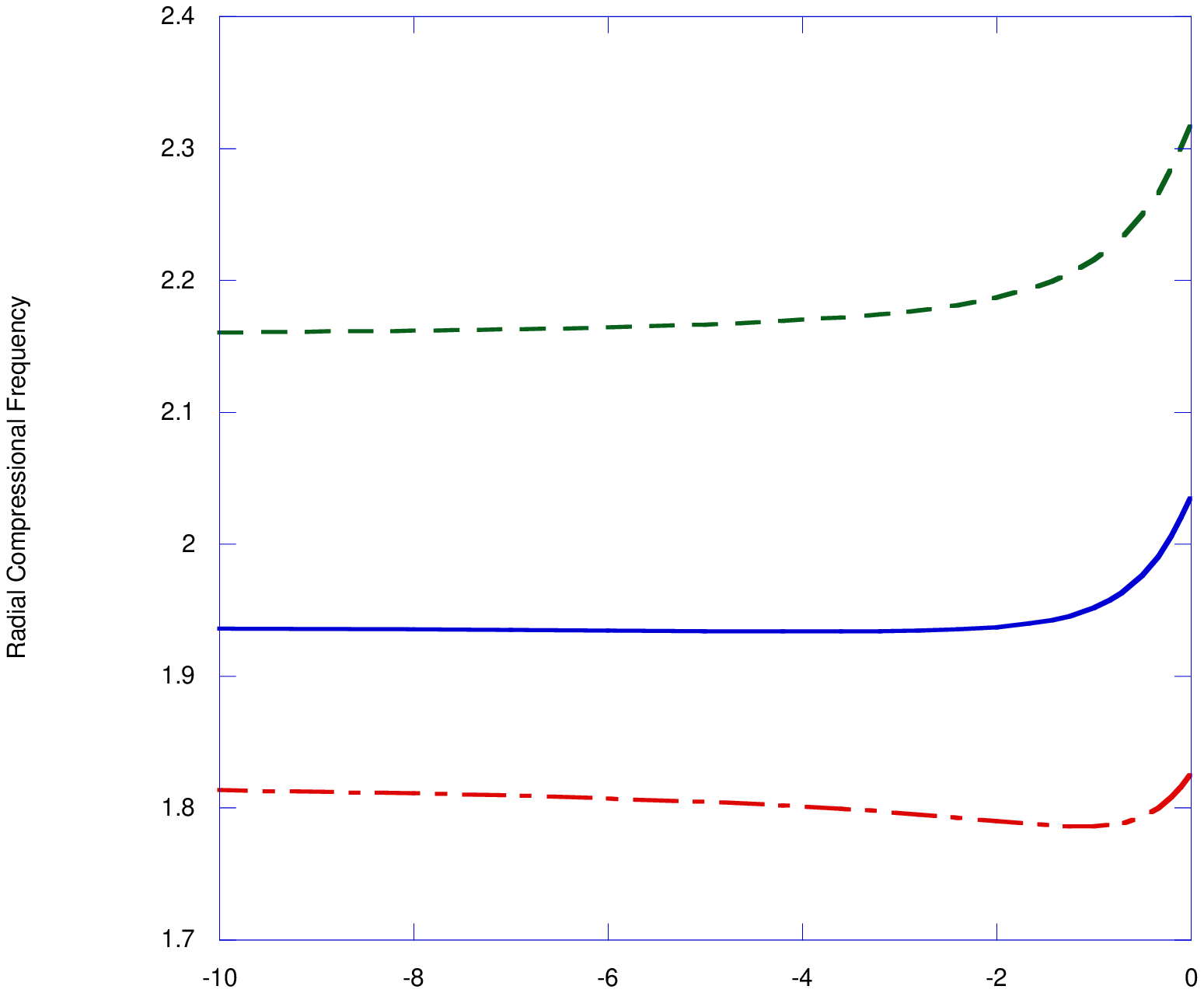}
\end{figure}
\begin{center}
$(N^{1/6}a/a_{ho})^{-1}$
\end{center}
Fig. 4.  Radial compressional frequency, $\omega_{rad}$, of the cloud of the $N=10^4$ fermions
(solid line) and $N=10^3$ fermions (dashed line)
 in unit of $\omega_{\perp}$ as a function
 of the dimensional parameter $(N^{1/6} a/a_{ho})^{-1}$. The trap parameter
$\lambda$ is assumed to be equal to $10^{-2}$. The lower line (dashed-dotted
 line)
 represents the results
in the   hydrodynamic
approximation, Eqs. (1) and (2), in which  $\omega_{rad}$ is independent of $N$ for a fixed $(N^{1/6} a/a_{ho})^{-1}$.

\pagebreak
\begin{figure}[ht]
\includegraphics{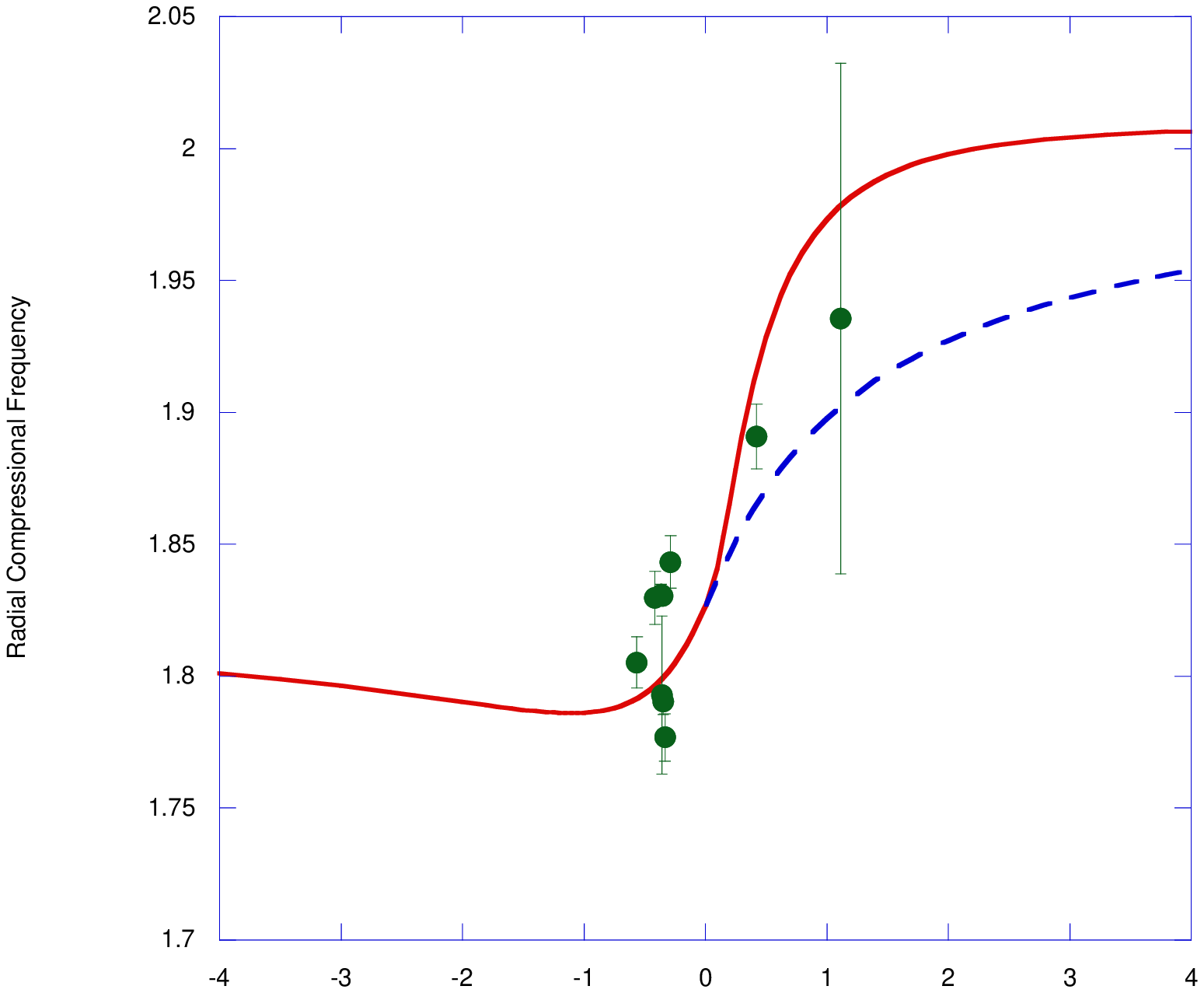}
\end{figure}
\begin{center}
$(N^{1/6}a/a_{ho})^{-1}$
\end{center}
Fig.5. The radial compressional frequency as  a function of $(N^{1/6}a/a_{ho})^{
-1}$. The solid line and the dashed line represent the results
calculated using the [2/2] Pad\'{e}
 approximation 
with $a_m=0.6 a$
and $a_m=2 a$, respectively. The solid circles with error bars are the
experimental results given by the Duke University group [1].

\pagebreak
{\bf References}
\vspace{8pt}

\noindent
1. J. Kinast, S. L. Hemmer, M. E. Gehm, A. Turlapov, and J. E. Thomas,
Phys. Rev. Lett. {\bf 92}, 150402 (2004). 

\noindent
2. C.A. Regal, M. Greiner, and D.S. Jin, Phys. Rev. Lett. {\bf 92}, 040403 (2004
).

\noindent
3. M. Greiner, C.A. Regal, and D.S. Jin, Nature {\bf 426}, 537 (2003).

\noindent
4. K.E. Stecker, G.B. Patridge, and R.G. Hulet, Phys. Rev. Lett. {\bf 91},
 080406 (2003).

\noindent
5. J. Cubizolles, T. Bourdel, S.J.J.M.F. Kokkelmans, G.V Shlyapnikov, and
 C. Salomon, Phys. Rev. Lett. {\bf 91}, 240401 (2003).

\noindent
6. S. Jochim, M. Bartenstein, A. Altmeyer, G. Hendl, C. Chin, J.H. Denschlag,
and
R. Grimm, Phys. Rev. Lett. {\bf 91}, 240402 (2003).

\noindent
7. S. Jochim, M. Bartenstein, A. Altmeyer, G. Hendl, S. Riedl, C. Chin,
J.H. Denschlag, and  R. Grimm, Science {\bf 302}, 2101 (2003).

\noindent
8. M.W. Zwierlein, C.A. Stan, C.H. Schunck, S.M.F. Raupach, S. Gupta,
Z. Hadzibabic, W. Ketterle, Phys. Rev. Lett. {\bf 91}, 250401 (2003).

\noindent
9. C.A. Regal and D.S. Jin, Phys. Rev. Lett. {\bf90}, 230404 (2003).

\noindent
10. K.M. O'Hara, S.L. Hemmer, M.E. Gehm, S.R. Granade, and  J.E. Thomas,
Science {298}, 2179 (2002).

\noindent
11. M. Bartenstein, A. Altmeyer, S. Riedl, S. Jochim, C. Chin, J.
Hecker Denschlag, and R. Grimm, Phys. Rev. Lett. {\bf92}, 120401 (2004).

\noindent
%12. L. Pitaevskii and S. Stringari, {\it  Bose-Einstein Condensation} (Clarendon Press,
% Oxford 2003).
12. Hui Hu, A. Minguzzi, Xia-Ji Liu, and M.P. Tosi, cond-mat/0404012.

\noindent
13. C. Menotti, P. Pedri and S. Stringari, Phys. Rev. Lett. 
{\bf 89}, 25042 (2002).

\noindent
14.  M. Cozzini and S. Stringari, Phys. Rev. Lett. {\bf 91},
 070401 (2003).

\noindent
15. S. Stringari, Europhys. Lett. {\bf 65}, 749 (2004).

\noindent
16. A. Bulgac and Y. Yu, Phys. Rev. Lett. {\bf 88}, 042504 (2002);
{\bf 91}, 190404 (2003); A. Bulgac, Phys. Rev. C{\bf 65}, 051305 (2002);
Y. Yu and Bulgac, Phys. Rev. Lett. {\bf 90}, 222501 (2003);
{\bf 90}, 101101 (2003).

\noindent
17. L.N. Oliveira, E.K.U. Gross and W. Kohn, Phys. Rev. Lett., {\bf 60}, 2430 (1988);
S. Kurth, M. Marques, M. L\"{u}ders, and E.K.U. Gross, Phys. Rev. Lett. {\bf 83}, 2628 (1999);
E.K.U. Gross, M. Marques, M. L\"{u}ders, and L. Fast, AIP Conf. Proc. {\bf 577}, 177 (2001). 

\noindent
18. D.A. Kirzhnitz, {\it Field Theoretical Methods in Many-Body Systems}
(Pergamon Press, London 1967).

\noindent
19. von C.F.  Weizs\"{a}cker, Z. Phys. {\bf 96}, 431 (1935).

\noindent
20. V.G. Kartavenko, K.A. Gridnev, J. Maruhn, and W. Greiner, Phys. At. Nucl.
{\bf 66}, 1439 (2003) and references therein;
P.K. Achorya, L.S. Bartolotti, S.B. Sears, and R.G. Parr, Proc. Natl. Acad. Sci. USA, {\bf 77}, 6978 (1980);
J.L. Gasquez and J. Rolles, J. Chem Phys. {\bf 76}, 1467 (1982).

\noindent
21. B.M. Deb and  S.K. Ghosh, Int. J. Quantum Chem. {\bf 23}, 1 (1983);
B.M. Deb and P.K. Chattaraj, Phys. Rev. A{\bf 39}, 1696 (1989);
B.M. Deb, P.K. Chattaraj, and  S. Mishra, Phys. Rev. A{\bf 43}, 1428 (1991); R. Singh
 and B.M. Deb, Phys. Rep. {\bf 311}, 47 (1999).

\noindent
22. The only difference from equations holding for bosons [23,24] is given by
density dependence of $\epsilon(n)$. We do not consider three-body
 recombinations, since these processes play an important role near p-wave
two-body Feshbach resonance [25].

\noindent
23. Y.E. Kim and A.L. Zubarev, Phys. Rev. A{\bf 67}, 015602 (2003).

\noindent
24. Y.E. Kim and A.L. Zubarev, Phys. Rev. A{\bf 69}, 023602 (2004).

\noindent
25. H. Suno, B.D. Esry and C.H. Greene, Phys. Rev. Lett. {\bf 90}, 053202 (2003)
.

\noindent
26. W. Lenz, Z. Phys. {\bf 56}, 778 (1929); K. Huang and C.N. Yang, 
Phys. Rev. {\bf105}, 767 (1957);
T.D. Lee and C.N. Yang, ibid {\bf 105}, 1119 (1957);
V.N. Efimov and M.Ya. Amus'ya, Zh. Eksp. Teor. Fiz. {\bf 47}, 581 (1964) [
Sov. Phys. JETP {\bf 20},388 (1965)].

\noindent
27.  G.A. Baker, Jr., Int. J. Mod. Phys. B{\bf15}, 1314 (2001); Phys. Rev. C{\bf
60},
054311 (1999).

\noindent
28. J. Carlson, S.-Y. Chang, V.R. Pandharipande, and K.E. Schmidt, 
Phys. Rev. Lett. {\bf
 91},
050401 (2003).

\noindent
29. A.J. Leggett, in {\it Modern Trends in the Theory of Condensed 
Matter},
 edited by A. Pekalski and R. Przystawa, Lecture Notes in Physics 
Vol. 115 (Springer-Verlag, Berlin, 1980) pp. 13-27;
P. Nozi\`{e}res and S. Schmitt-Rink, J. Low. Temp. Phys. {\bf 59}, 195 (1985).

\noindent
30. J.R. Engelbrecht, M. Randeria, and C.A.R. Sa de Melo, Phys. Rev. B{\bf 55},
15153 (1997).

\noindent
31. D.S. Petrov, C. Salomon and G.V. Shlyapnikov, cond-mat/0309010.

\noindent
32. P. Pieri and G.S. Strinati, Phys. Rev. B{\bf 61}, 15370 (2000).

\noindent
33. Y.E. Kim and A.L. Zubarev, cond/mat/0403085; Phys. Lett. A (in print).

\noindent
34. V.R. Pandharipande, Nucl. Phys. A{\bf 174}, 641 (1971);
S. Cowel, H. Heiselberg, I.E. Morales, V.R. Pandharipande, and C.J. Pethick,
 Phys. Rev. Lett. {\bf 88}, 210403 (2002); H. Heiselberg, J. Phys.  B{\bf 37},
S141 (2004); H. Heiselberg, cond-mat/0403041.

\noindent
35. To calculate the ground-state density we have used a highly accurate
 variational approach of Ref.[36]. This method gives, for example, 
  in unitary limit for the case of very large $N$ the value of energy
$E/(N^{4/3} \lambda^{1/3} (1+\beta)^{1/2})=1.08486$, which is very close to 
the exact value $3^{4/3}/4\approx 1.08169$ (relative error is less than 0.3\%).

\noindent
36. M.P. Singh and A.L. Satheesha, Eur. Phys. J., D{\bf 7}, 321 (1998).

\noindent
37. We have used the data from Ref.[38] to convert $a$ to $B$. We note here
that in general  a Feshbach resonance may lead to the density dependence of the
effective interaction
  (for bosons cases see, for example, [39,40]).
In Ref.[41], the resonant position was accurately determined to be $822\pm 3$G. 

\noindent
38. K. M. O'Hara, S. L. Hemmer, S. R. Granade, M. E. Gehm, J. E. Thomas,
     V. Venturi, E. Tiesinga, and C. J. Williams, Phys. Rev. A{\bf 66},
 041401 (2002).

\noindent
39. Y.E. Kim and A.L. Zubarev, Phys. Lett. A{\bf 312}, 277 (2003).

\noindent
40. V.A. Yurovsky, cond-mat/0308465.

\noindent
41. M. W. Zwierlein, C. A. Stan, C. H. Schunck, S. M. F. Raupach, 
A. J. Kerman, and W. Ketterle, Phys. Rev. Lett. {\bf 92}, 120403 (2004).

\noindent 
42. A. Recati, P.O. Fedichev, W. Zwerger, and P. Zoller,
Phys. Rev. Lett. {\bf 90}, 020401 (2003);
 G.E. Astrakharchik, D. Blume, S. Giorgini, and L.P. Pitaevskii,
 cond/mat/0312538.

\noindent
43. S.J. Puglia, A. Bhattacharyya, and R.J. Furnstahl, Nucl. Phys. A{\bf 723},
145 (2003).

\noindent
44. M. Bartenstein, A. Altmeyer, S. Riedl, S. Jochim, C. Chin,
 J. Hecker Denschlag, and R. Grimm,  Phys. Rev. Lett. {\bf 92}, 203201 (2004).

\noindent
45. R.G. Parr and W. Yang, {\it Density Functional Theory of Atoms and 
Molecules}, (Oxford Univ. Press, New York, 1989).

\noindent
46. G.E. Astrakharchik, J. Boronat, J. Casulleras, and S. Giorgini,
 cond/mat/0406113;
S.Y. Chang, V.R. Pandharipande, J. Carlson and K.E. Schmidt, physics/0404115.
\end{document}